\documentclass[conference]{IEEEtran}

\usepackage{ifpdf}
\usepackage{cite}
\usepackage{multirow}
\usepackage{bm}
\usepackage{bbm}
\usepackage{booktabs}
\usepackage{color}
\usepackage{subfigure}

 \usepackage{graphicx,xcolor}

\usepackage[cmex10]{amsmath}
\usepackage{url}

\usepackage{amsmath}
\usepackage{amssymb}
\usepackage{amsfonts}
\usepackage{mathtools}
\usepackage{amsthm}

\theoremstyle{definition}
\theoremstyle{definition}

\usepackage{algorithm,algpseudocode}
\algdef{SE}[DOWHILE]{Do}{DoWhile}{\algorithmicdo}[1]{\algorithmicwhile\ #1}
\algdef{SE}[SUBALG]{Indent}{EndIndent}{}{\algorithmicend\ }%
\algtext*{Indent}
\algtext*{EndIndent}

\usepackage{tikz} 
\usetikzlibrary{arrows,shapes,backgrounds,fit}

\begin{document}
\title{Cooperative Sensing in Deep RL-Based \\Image-to-Decision Proactive Handover\\
for mmWave Networks}

\author{
	\IEEEauthorblockN{
		\normalsize Yusuke Koda\IEEEauthorrefmark{1}\IEEEauthorrefmark{2},
		\normalsize Koji Yamamoto\IEEEauthorrefmark{1}\IEEEauthorrefmark{3},
		\normalsize Takayuki Nishio\IEEEauthorrefmark{1}, and
		\normalsize Masahiro Morikura\IEEEauthorrefmark{1}
	}
	\IEEEauthorblockA{
		\IEEEauthorrefmark{1}\small Graduate School of Informatics, Kyoto University,
		Yoshida-honmachi, Sakyo-ku, Kyoto 606-8501, Japan
	}
	\IEEEauthorblockA{
		\IEEEauthorrefmark{2}
		koda@imc.cce.i.kyoto-u.ac.jp
		\IEEEauthorrefmark{3}kyamamot@i.kyoto-u.ac.jp
	}
}

\maketitle
\begin{abstract}
	For reliable millimeter-wave (mmWave) networks, this paper proposes cooperative sensing with multi-camera operation in an image-to-decision proactive handover framework that directly maps images to a handover decision.
	In the framework, camera images are utilized to allow for the prediction of blockage effects in a mmWave link, whereby a network controller triggers a handover in a proactive fashion. 
	Furthermore, direct mapping allows for the scalability of the number of pedestrians.
	This paper experimentally investigates the feasibility of adopting cooperative sensing with multiple cameras that can compensate for one another's blind spots.
	The optimal mapping is learned via deep reinforcement learning to resolve the high dimensionality of images from multiple cameras.
	An evaluation based on experimentally obtained images and received powers verifies that a mapping that enhances channel capacity can be learned in a multi-camera operation.
	The results indicate that our proposed framework with multi-camera operation outperforms a conventional framework with single-camera operation in terms of the average capacity.
\end{abstract}
\IEEEpeerreviewmaketitle


%





%
\section{Introduction}
\label{sec:intro}
Millimeter-wave (mmWave) communication is a key technology for fifth-generation (5G) wireless systems\cite{agiwal2016next}.
The large bandwidth available at mmWave frequencies
 enables transmission of multiple gigabits of data per second.
With mmWave communication, 5G wireless systems can support contents that require high data rates, such as ultrahigh definition and three-dimensional video contents\cite{agiwal2016next}.
However, mmWave communication undergoes greater degradation than microwave communications in terms of received power because of blockage effects caused by pedestrians \cite{koda_measurement, haneda2015channel, maccartney2017flexible}.
The blockage effects degrade the performance of mmWave communication and prevent support for contents that require higher data rates.

Handover among multiple base stations (BSs) is a promising approach to maintain the performance of a mmWave link. 
Many studies have investigated the system design of the handover mechanism\cite{multiAPs, sun2017reinforcement, mmwave_mdp}.
In the mechanisms, a handover is triggered through a blockage prediction based on such qualities of the relevant link as channel state information, received power, and throughput.
Therefore, when the variation in the link quality is so fast that the blockage effect cannot be predicted, handover systems cannot avoid degradation in the qualities of mmWave links.

To avoid the degradation in the quality of the mmWave link, we have in past research proposed a camera-assisted proactive handover framework\cite{proactive, proactive3}. 
The framework predicts blockages using camera images and triggers handovers in a proactive fashion.
The research in \cite{proactive, proactive3} has conceptualized the camera-assisted proactive handover and shown that using camera images, a handover can be performed approximately one second before blockage occurs through proof-of-concept prototyping.

In the camera-assisted proactive handover framework, it remains challenging to optimize the rules for a handover decision because of the high dimensionality of the camera images.
To combat the high dimensionality, in previous studies\cite{proactive, proactive3, koda_infocom}, a network controller estimates lower-dimensional information related to pedestrian movement, such as positions/velocities, using a human tracking system, and makes handover decisions based on the estimated information.
However, the solutions limit the number of pedestrians and cannot deal with situations where more pedestrians cause blockage effects.
For example, \cite{koda_infocom} assumed that a single pedestrian causes blockage effects.
Thus, this solution cannot deal with a situation where two or more pedestrians cause blockage effects.

To construct the optimal decision rule based on information containing high dimensionality, deep reinforcement learning (RL) provides a new approach\cite{DQN2}.
Deep RL enables the direct mapping of high-dimensional sensory input to an optimal decision without explicitly estimating lower-dimensional information necessary for control.
Because of the direct mapping, deep RL is applicable to more generic tasks rather than specific ones.
The study \cite{DQN2} showed that deep RL can successfully map high-dimensional image inputs to an action in Atari 2600 games, and surpasses the performance of a professional human player in many games.
Although deep RL has been applied to some research areas\cite{wang2018deep, wan2017reinforcement, xu2017deep}, 
its application to the problem of constructing the optimal decision rule in a camera-assisted proactive handover is a new direction.

Recently, we introduced a paradigm called the deep RL-based image-to-decision proactive handover (I2D-PH) framework\cite{koda_tvt_arxiv}.
The framework directly maps camera images\footnote{We used depth images\cite{depth} whose pixels measured the distance between obstacles and a camera. Depth images allow us to obtain geometric relations between the components of a scene. In the following discussion, we assume that depth images are available to a network controller.} to a handover decision without explicitly estimating the positions/velocities of pedestrians.
The direct mapping allows for the scalability of the number of pedestrians, i.e., it can deal with situations where an arbitrary number of pedestrians cause blockage effects as long as they are within the range of coverage of the camera.
We experimentally demonstrated that via deep RL, the optimal mapping can be learned with the objective of enhancing overall performance\cite{koda_tvt_arxiv}.
However, blockage effects caused by a pedestrian out of the camera's coverage cannot be avoided.
Thus, to deal with blockage effects caused by an arbitrary number of pedestrians, multiple cameras should be used.
There is room to investigate the feasibility of using multi-camera operation in the I2D-PH framework.

This paper investigates the feasibility of adopting cooperative sensing with multi-camera operation to the I2D-PH framework.
The multiple cameras allow blockage prediction while compensating for one another's blind spots.
Deep RL is developed to learn the optimal mapping from multiple images to a handover decision.

The main contributions of this paper are summarized as follows:
\begin{itemize}
	\item To successfully adopt cooperative sensing with multi-camera operation to the I2D-PH framework, we design a decision process where the state includes multiple images from multiple cameras.
	\item We demonstrate the feasibility of adopting cooperative sensing to the I2D-PH framework through an evaluation using experimentally obtained camera images and received powers.
	The evaluation validates that the framework with multi-camera operation outperforms a conventional framework with single-camera operation in terms of the average capacity.
\end{itemize}

The remainder of this paper is organized as follows.
Section~\ref{sec:system_model} provides the system model of the I2D-PH framework with cooperative cameras.
Section~\ref{sec:formulation} discusses the RL formulation for the handover, and Section~\ref{sec:experiment} provides an experimental evaluation of the proposed handover framework.
Finally, Section~\ref{sec:conclusion} provides the conclusions of the paper.
\section{I2D-PH Framework with Cooperative Sensing}
\label{sec:system_model}

\subsection{System Model}
\begin{figure}[!t]
	\centering
	\includegraphics[width=0.9\columnwidth]{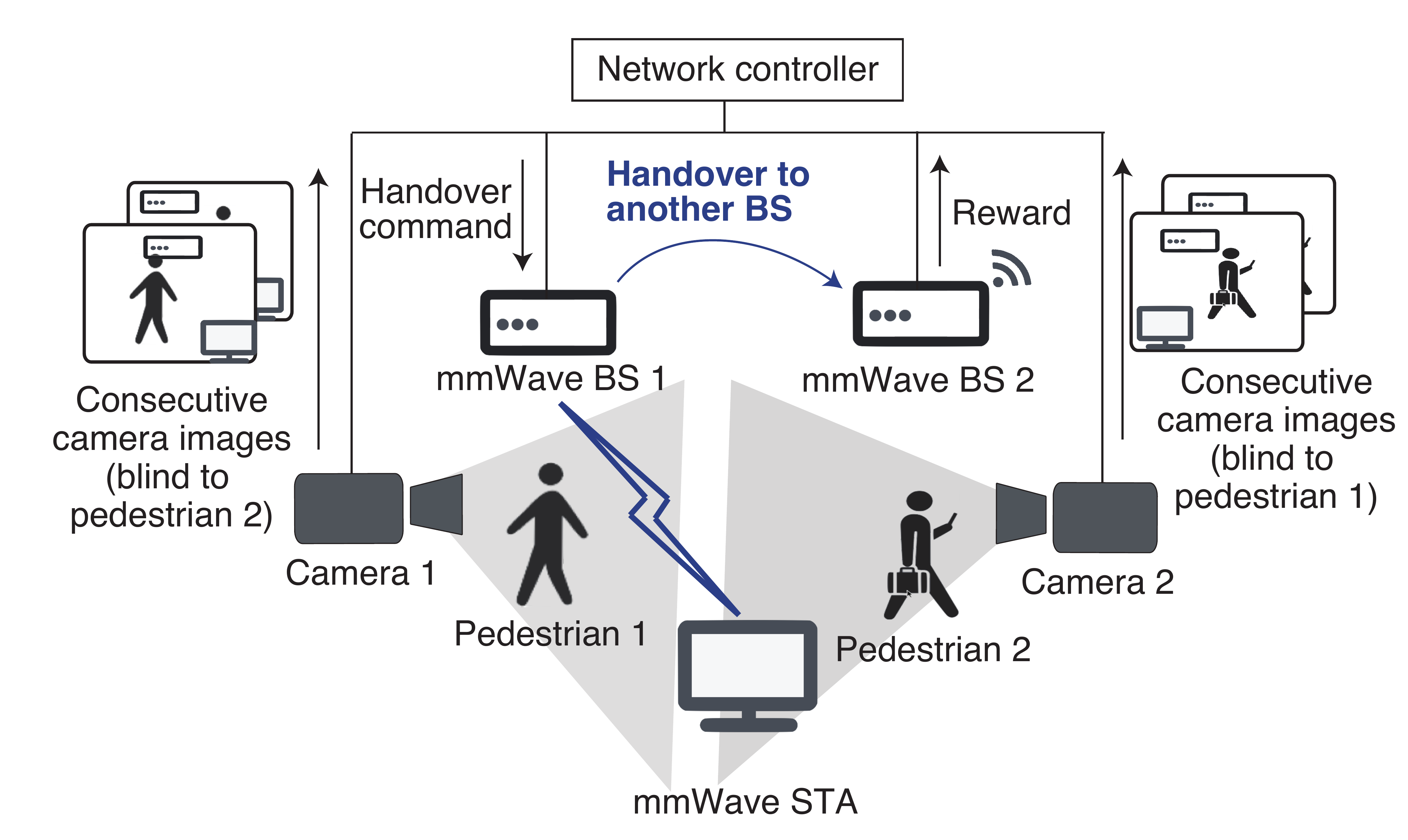}
	\caption{I2D-PH framework with cooperative camera sensing. Camera~1 compensates for the blind spots of camera~2, and vice versa.}
	\label{fig:system_model}
\end{figure}

We consider a scenario where multiple mmWave BSs, a station (STA), and multiple cameras are deployed as shown in Fig.~\ref{fig:system_model}\footnote{We assume that the positions of the STA and BSs are quasi-static, i.e., the variation in the positions occurs over a longer time scale than the learning procedure.
The assumption is due to our focus on solving link blockage problems caused by moving pedestrians.
The consideration of the variation in the positions of the STA and BSs during the learning procedure is beyond the scope of this study.}.
The STA is associated with one of the BSs, and can perform handover to another BS according to a handover command from a network controller.

Note that we consider multi-camera operation where the cameras compensate for one another's blind spots.
Fig.~\ref{fig:system_model} shows an example, where camera~1 is blind to pedestrian~2 because of limited coverage.
Similarly, camera~2 is blind to pedestrian~1.
By combining images from the cameras, the network controller can be aware of the movement of all pedestrians.
Based on the combined images, the network controller predicts blockage effects and decides whether a handover should be triggered.

We also consider whether a handover should be triggered with regard to its cost\cite{koda_tvt_arxiv}.
The communication between the BS and the STA can be disrupted because of the necessary procedure for the association\cite{handover_cost}.
We define the duration for which the communication is disrupted as service disruption time $T_{\rm dis}$.

\subsection{Direct Mapping of Camera Images to Handover Decisions}
The network controller directly maps consecutive images from multiple cameras to a handover decision without explicitly estimating the positions/velocities of the pedestrians from the images.
That is, the network controller triggers handovers based on the image pixels and variations in them.
Because they reflect the positions/velocities of each pedestrian in the images, the framework can deal with the blockage effects caused by each pedestrian within the images irrespective of their number.

\subsection{Learning Procedure for Optimal Mapping}
The network controller learns the optimal mapping of consecutive images from each camera to handover decisions via deep RL.
As shown in Fig.~\ref{fig:system_model}, the network controller obtains images from each camera, triggers a handover in a trial-and-error fashion, and subsequently obtains a reward---a performance metric in the mmWave link such as received power, throughput, or data rate---from the BS associated with the STA.
Based on the history of the camera images, the handover decision, and the reward, the network controller learns the optimal mapping that maximizes the expected cumulative sum of rewards.
The learning procedure continues for a predefined duration.

\section{Model Formulation}
\label{sec:formulation}
\subsection{Reinforcement Learning}
General RL algorithms including deep RL are executed in a Markov decision process (MDP).
An MDP consists of the following four elements: a state set $\mathcal{S}$, an action set $\mathcal{A}$, a reward function $r:\mathcal{S}\times\mathcal{A}\times\mathcal{S}\to\mathbb{R}$, and transition probabilities $q:\mathcal{S}\times\mathcal{A}\to \Omega(\mathcal{S})$, where $\Omega(\mathcal{S})$ denotes the collection of probability distributions over $\mathcal{S}$.
At each decision epoch $t\in\mathbb{N}$, a decision maker observes state information $s_t\in\mathcal{S}$.
Subsequently, the decision maker selects an action on the basis of the {\it policy} $\pi:\mathcal{S}\to\mathcal{A}(s_t)$, where $A(s_t)\subseteq \mathcal{A}$ denotes the set of possible actions when state $s_t$ is observed.
Given the current state $s_t$ and selected action $a_t\in\mathcal{A}(s_t)$, the state transitions to $s_{t + 1}\in\mathcal{S}$ in the next decision epoch $t + 1$ according to transition probability $q(s_{t + 1}, s_t, a_t)$; thereafter, the decision maker is given a reward $r(s_{t + 1}, a_{t}, s_t)$.

The objective of RL is to learn the optimal policy $\pi^{\star}$ that maximizes the expected cumulative sum of future rewards. 
Let $V^{\pi}(s)$ denote the expectation under policy $\pi$ and the initial state $s$, i.e., $V^{\pi}(s) \coloneqq \mathbb{E}\!\left[\,\sum_{t' = 0}^{\infty}\gamma^{t'}r\bigl(s_{t + t' + 1}, \pi(s_{t + t'}), s_{t + t'}\bigr)\,\middle|\,s_t = s\right]$, 
where $\gamma\in[0, 1)$ denotes the discount factor.
The optimal policy $\pi^{\star}$ satisfies the following condition:
\begin{align}
	\label{eq:optimal_policy}
	V^{\pi^{\star}}(s)\geq V^{\pi}(s),\quad \forall s\in\mathcal{S}, \forall \pi.
\end{align}
In the MDP wherein $\mathcal{S}$ and $\mathcal{A}$ are both countable non-empty sets, there exists at least one optimal policy\cite{sutton}.

To obtain the optimal policy, it is sufficient to obtain the optimal action-value function $Q^{\star}:\mathcal{S}\times\mathcal{A}\to\mathbb{R}$.
The optimal action-value function is defined as follows:
\begin{multline}
	\label{eq:opt_q_func}
	Q^{\star}(s, a) \coloneqq \mathbb{E}_{s'}\!\left[r(s', a, s) + \gamma V^{\pi^\star}(s')\,\vert\,s, a\right],\\ s\in \mathcal{S}, a\in\mathcal{A}(s),
\end{multline}
where $\mathbb{E}_{s'}[\,\cdot\,\vert\, s, a\,]$ denotes the expectation operator under the transition probability $q(s', s, a)$.
This is attributed to the fact that the policy that selects the action that maximizes $Q^{\star}(s, a)$ is optimal.
In this paper, the optimal action-value function is learned via deep RL\cite{DQN2}\footnote{We apply the algorithm in\cite{DQN2} with a neural network (NN) architecture in \cite{koda_tvt_arxiv}; thus, we do not discuss the deep RL algorithm and NN architecture due to  restrictions of space.}.

\subsection{Decision Process in I2D-PH Framework with Multi-Camera Operation}
\label{subsubsec:definition}
We formulate the decision process where the network controller makes handover decisions in I2D-PH framework with multi-camera operation.
We define the states, actions, rewards, and the state transition rule as follows.

\subsubsection{States}
	For the network controller to utilize images from multiple cameras to make handover decisions, we design the states such that they contain consecutive images from each camera.
	Let the numbers of deployed cameras and consecutive images be denoted by $I$ and $N$, respectively.
	We set the state set as follows:
		  \begin{align}
			\label{eq:state_set}
		      \mathcal{S}\coloneqq \underbrace{\mathcal{X}^{(1)}\times\dots\times\mathcal{X}^{(1)}}_{N}\times\dots\times\underbrace{\mathcal{X}^{(I)}\times\dots\times\mathcal{X}^{(I)}}_{N}\times\mathcal{J}\times\mathcal{C}.
		  \end{align}
	In \eqref{eq:state_set}, $\mathcal{X}^{(i)}$ for $i = 1, 2, \dots, I$ denotes the set of all possible images from the $i$th camera, $\mathcal{J}\coloneqq\{1,\dots,J\}$ denotes the set of BS indices, and $ \mathcal{C} \coloneqq \{\,c\mid c\in\mathbb{Z},\  0\leq c\leq \lfloor T_{\mathrm{dis}}/\tau \rfloor\,\}$ denotes the set of remaining decision epochs until the service disruption time ends, where $J$ denotes the number of  deployed BSs, $\lfloor \cdot \rfloor:\mathbb{R}\to\mathbb{R}$ denotes the floor function, and $\tau$ denotes the interval between successive decision epochs.
	
	Let $s_t = \bigl(x^{(1)}_{t},\dots, x^{(1)}_{t - N + 1},\dots,x^{(I)}_t, \dots, x^{(I)}_{t - N + 1}, j_t, c_t\bigr)\in\mathcal{S}$ denote the state at the decision epoch $t\in \{\,t\mid\,t\in\mathbb{Z},\  t\geq N\}$.
	The element $x^{(i)}_{t - k}\in\mathcal{X}^{(i)}$ for $i\in\{1, \dots, I\}$ and $k\in \{0, 1, \dots, N - 1\}$ is set as the image observed at the decision epoch $t - k$ in the $i$th camera.
	The element $j_t\in\mathcal{J}$ is set as the index of the BS associated with the STA.	  
	The element $c_t\in \mathcal{C}$ is set as the number of remaining decision epochs that the network controller experiences until service disruption ends.
	When the decision epoch is not within the service disruption time, $c_t$ is set to zero.
	
	\subsubsection{Actions} 
	      We set the set of possible actions $\mathcal{A}(s_t)$ be as follows:
	      \begin{align}
		      \mathcal{A}(s_t)\coloneqq
		      \begin{cases}
			      \mathcal{J}, & c_t = 0;    \\
			      \{j_t\},       & c_t \neq 0.
		      \end{cases}
	      \end{align}
	      That is, the controller selects the index of a BS when the decision epoch is not within the service disruption time; otherwise, the controller selects the index of the BS to which a handover is triggered.

	\subsubsection{Reward}
	 We set the reward as a performance metric in the link provided by the BS associated at the given time with the STA:
	      \begin{align}
		      \label{eq:reward}
		      r(s_{t + 1}, a_t, s_t) \coloneqq
		      \begin{cases}
			      R^{(j_{t + 1})}_{t + 1}, & c_{t + 1} = 0;    \\
			      0,      & c_{t + 1} \neq 0,
		      \end{cases}
	      \end{align}
		  where $R^{(j_{t + 1})}_{t + 1}$ denotes the performance metric in the link provided by BS $j_{t + 1}$ at $t + 1$.
			In the performance evaluation, we set $R^{(j_{t + 1})}_{{t + 1}}$ as the capacity in the link provided by BS $j_{t + 1}$ as discussed in Section~\ref{sec:experiment}.
			In \eqref{eq:reward}, we set the reward to zero when the decision epoch $t + 1$ is within the service disruption duration\cite{handover_cost}.

			\subsubsection{State Transition}
			The state transition to the next state is as follows:
			Evidently, the consecutive images $\bigl(x^{(1)}_{t + 1}, \dots, x^{(1)}_{t - N + 2}, \dots, x^{(I)}_{t + 1}, \dots, x^{(I)}_{t - N + 2}\bigr)$ at $t + 1$ are determined by concatenating the images $x^{(1)}_{t + 1}, \dots, x^{(I)}_{t + 1}$ with the images in $s_t$ and removing the oldest images $x^{(1)}_{t - N + 1}, \dots, x^{(I)}_{t - N + 1}$.
			Based on the definition of the state, the term $j_{t + 1}$ is determined as follows:
			\begin{align}
				\label{eq:j_t_transition}
				j_{t + 1} = a_t.
			\end{align}
			The term $c_{t + 1}$ is determined as follows:
			\begin{align}
				\label{eq:c_t_transition}
				c_{t + 1} =
				\begin{cases}
					c_t - 1,                          & \text{$c_t \neq 0$};       \\
					\lfloor T_{\rm dis}/\tau \rfloor, & \text{$c_t = 0, a_t\neq j_t$}; \\
					0,                              & \text{$c_t = 0, a_t = j_t$}.
				\end{cases}
			\end{align}


\section{Performance Evaluation}
\label{sec:experiment}

\begin{figure}[!t]
	\centering
	\includegraphics[width = 0.9\columnwidth]{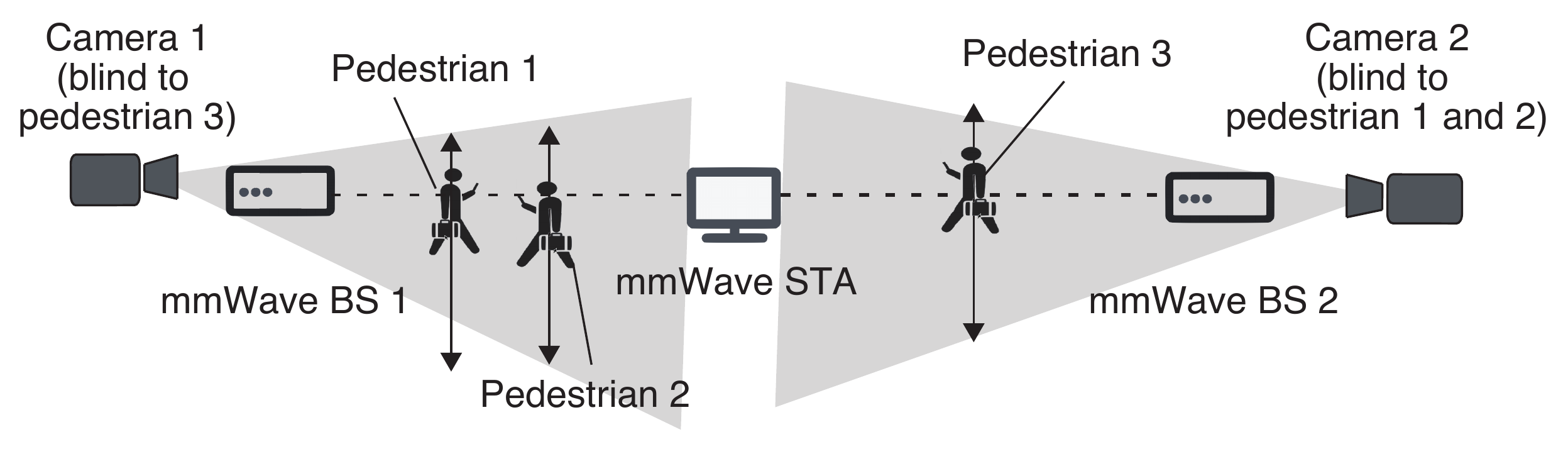}
	\caption{Considered mmWave links.}
	\label{fig:simulated}
\end{figure}

\begin{figure}[!t]
	\centering
	\subfigure[Measurement on BS~1 side.]{\includegraphics[width = 0.49\columnwidth]{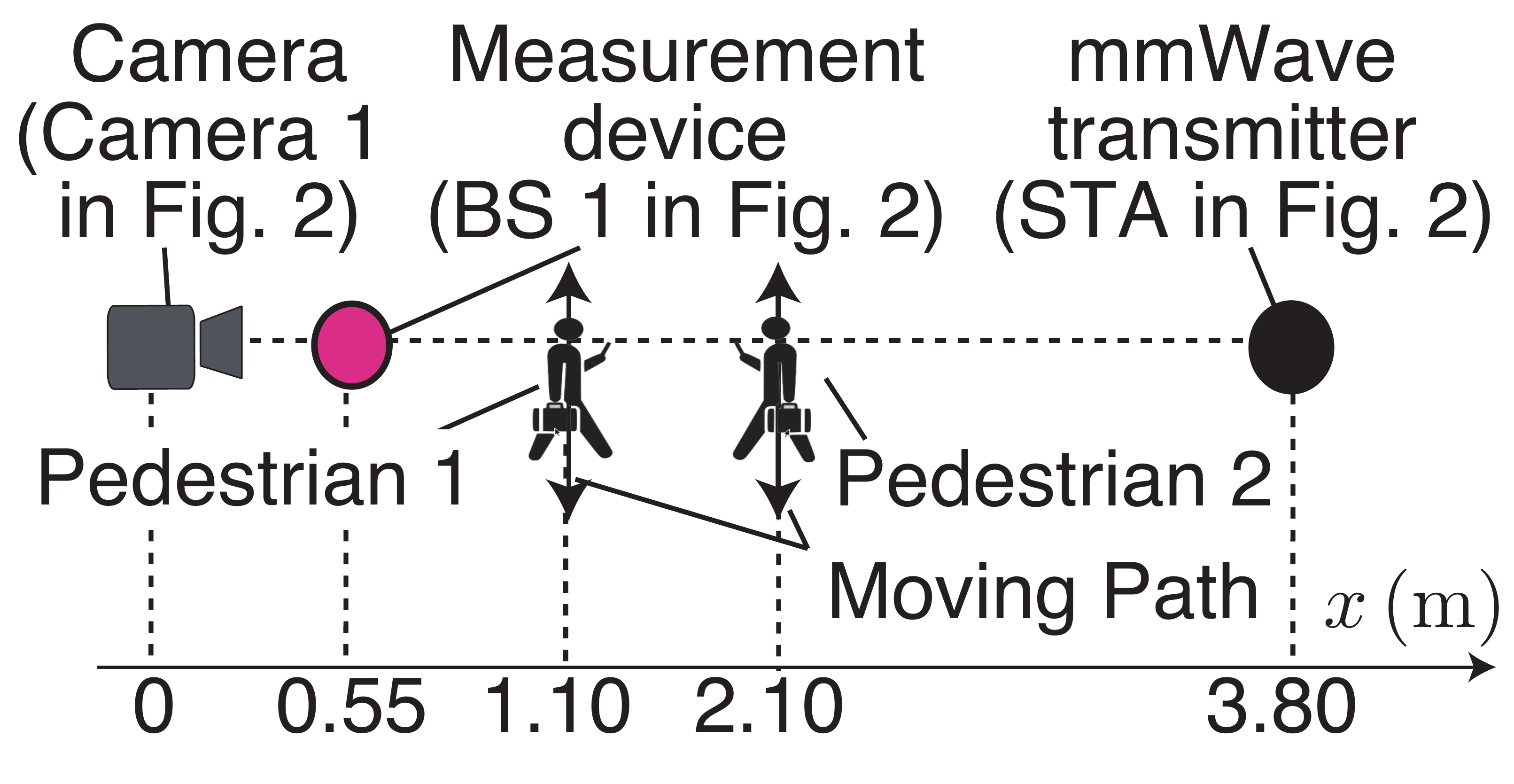}}\
	\subfigure[Measurement on BS~2 side.]{\includegraphics[width = 0.49\columnwidth]{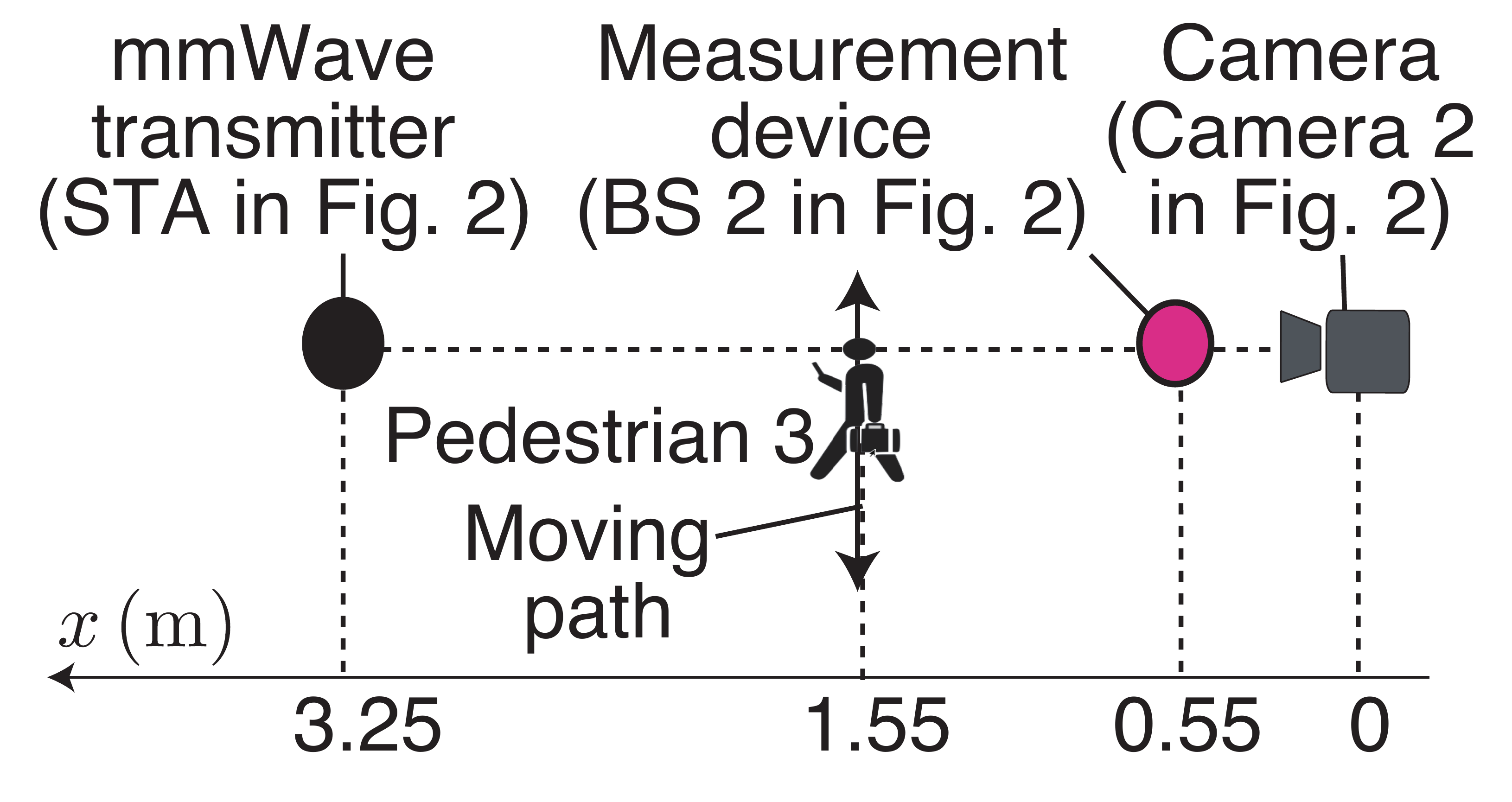}}
	\caption{Top view of the measurement environment.
	The measurement device, camera, and mmWave transmitter are correspond to BS~1/BS~2, camera~1/camera~2, and the STA in Fig.~\ref{fig:simulated}, respectively. Due to experimental limitations, we used the same measurement equipment in the two measurements.}
	\label{fig:mes}
\end{figure}

\begin{figure}[!t]
	\centering
	\subfigure[Picure of measurement environment.]{\includegraphics[width = 0.3\columnwidth]{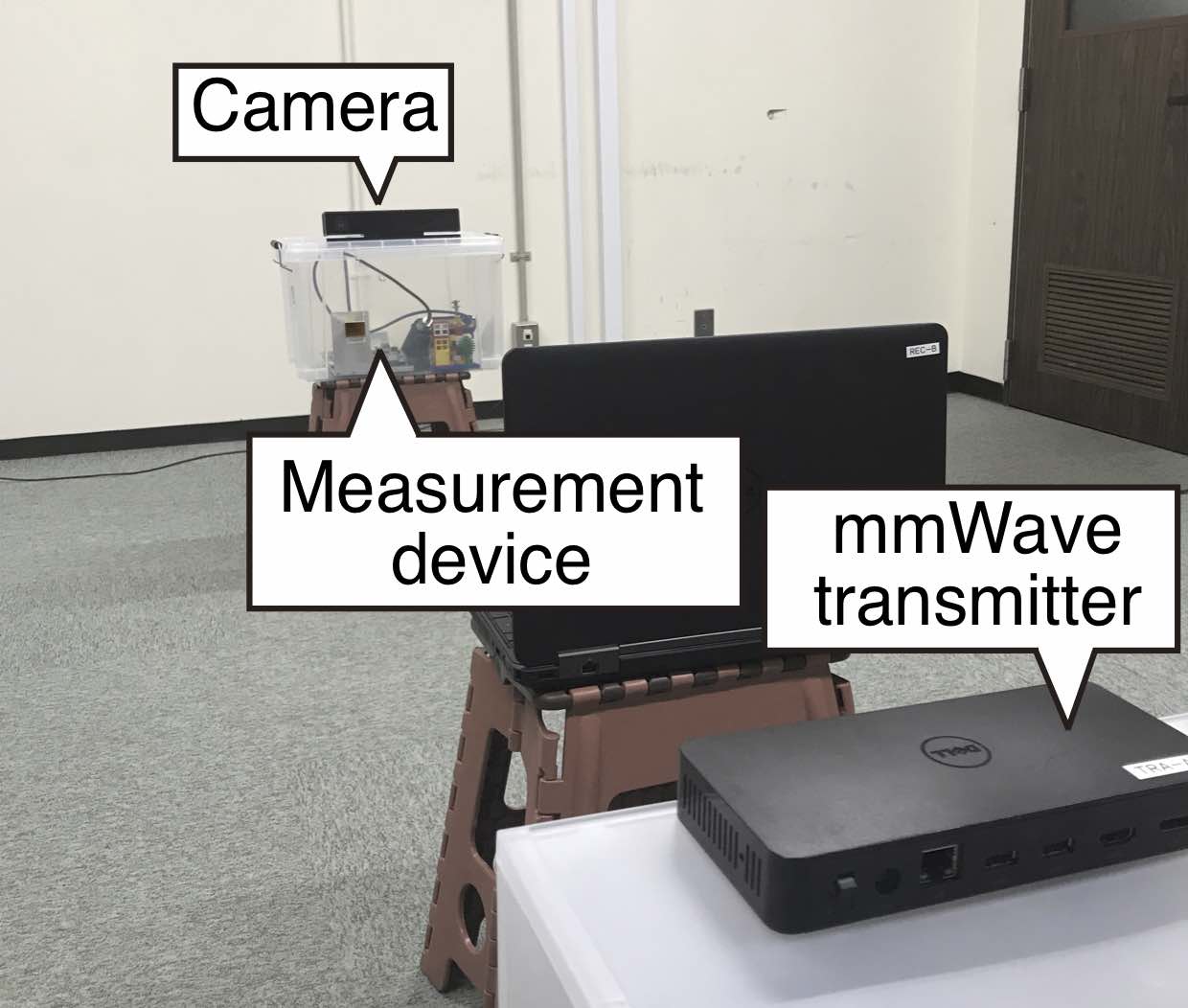}}
	\subfigure[Camera image on BS~1 side.]{\includegraphics[width = 0.3\columnwidth]{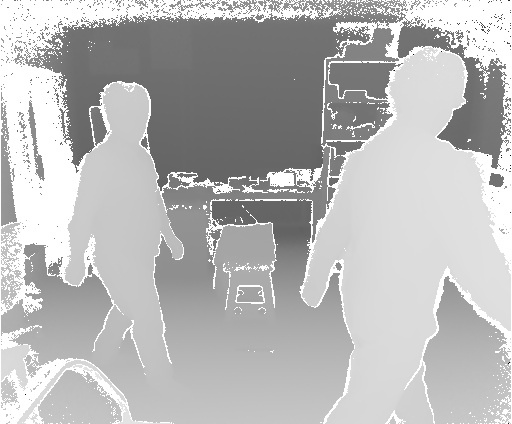}}
	\subfigure[Camera image on BS~2 side.]{\includegraphics[width = 0.33\columnwidth]{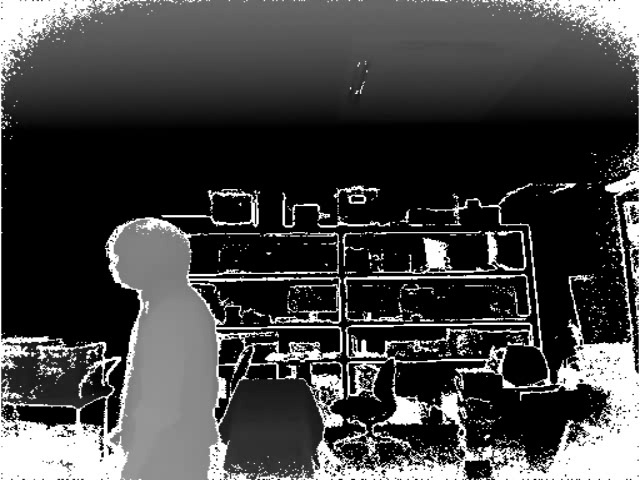}}
	\caption{Picture of measurement environment and images obtained in each measurement}
	\label{fig:mes_pic}
\end{figure}

\subsection{Evaluated Scenario}
\label{secV:subsec:evaluated_scenario}
We considered a scenario in which two BSs and an STA were deployed as shown in Fig.~\ref{fig:simulated}.
The STA was initially associated with the BS that observed higher received power compared with that of the counterpart when there were no pedestrians.
We called the BS initially associated with the STA BS~1 and the other BS~2.
BS~2 is a candidate BS in the case the link between BS~1 and the STA is blocked by pedestrians.

\subsection{Measurement Setup}
\label{secV:subsec:experiment_setup}
We conducted two measurements and obtained two data sets each of which contains received powers and camera images.
The former images were regarded as images from camera~1, while the latter were regarded as those from camera~2 in simulating the decision process as discussed in the following subsection.
The received powers obtained in the first measurement were used for calculating the reward at BS~1 while those obtained in the second measurement are used for calculating that at BS~2.

We deployed a mmWave transmitter, a measurement device, and a camera as shown in Fig.~\ref{fig:mes}.
The transmitter was considered to be the STA throughout the measurements.
The measurement device and camera were considered as BS~1 and camera~1, respectively, in the first measurement, while in the second measurement, they were considered BS~2 and camera~2, respectively.
The mmWave transmitter transmitted signals at a carrier frequency of 60.48\,GHz and subsequently, the measurement device measured the power of a part of the signals\cite{koda_measurement}. 
The transmitted signals were considered as uplink signals from the STA to BS~1/BS~2.
In the first measurement, two pedestrians walked along the moving path in Fig.~\ref{fig:mes}(a) and blocked the path between the transmitter and the measurement device.
Similarly, in the second measurement, a pedestrian walked along the moving path in Fig.~\ref{fig:mes}(b).
Table \ref{table:impl} summarizes the experimental equipment and parameters associated with the experiment.

\begin{table}[!t]
	\caption{Experimental Equipment and Parameters}
	\label{table:impl}
	\begin{center}
		\begin{tabular}{cc}	\toprule
			mmWave transmitter & Dell Wireless Dock D5000                  \\
			Spectrum analyzer  & Tektronix RSA306                          \\
			Down-converter     & Sivers IMA FC2221V                        \\
			Antenna            & Sivers IMA Horn antenna, 24\,dBi          \\
			Depth camera       & Microsoft Kinect  \\
			& for Windows (Model:1656)\\
			Channel                    & 60.48\,GHz           \\
			Sampling frequency         & 56\,MHz              \\
			Transmit antenna gain      & 10\,dBi \cite{d5000} \\
			Receive antenna gain       & 24\,dBi              \\
			Measurement bandwidth $W_1$, $W_2$ & 40\,MHz, 20\,MHz    \\
			\bottomrule
		\end{tabular}
	\end{center}
\end{table}



\subsection{Simulation Procedure of Decision Process}
\label{secV:subsec:simulate_MDP}
We divided the camera images and received powers into two parts to perform learning and performance evaluation based on different sets of data.
Let the camera images and received powers obtained in the $i$th measurement be denoted by $\bigl(x^{(i)}_t\bigr)_{t\in\mathcal{T}}$ and $\bigl(P^{(i)}_t\bigr)_{t\in\mathcal{T}}$, respectively, where $x^{(i)}_t$ denotes the $t$th image obtained in the $i$th measurement, $P^{(i)}_t$ denotes the received power obtained at the same time, and $\mathcal{T} = \{1, 2, \dots, T\}$ denotes the set of time indices.
We divided $\mathcal{T}$ into the following two subsets: $\mathcal{T}_1 = \{1, 2, \dots, T'\}$ and $\mathcal{T}_2 = \{T' + 1, T' + 2, \dots, T\}$, where $1 < T' < T$.

We simulated the decision process in the learning procedure using $\bigl(x^{(i)}_t\bigr)_{t\in \mathcal{T}_1}$ and $\bigl(P^{(i)}_{t}\bigr)_{t\in\mathcal{T}_1}$ for $i \in \{1, 2\}$.
The decision epoch was set as the time step in which an image was obtained.
The decision process started at the time step at which $x^{(1)}_{N}$ and $x^{(2)}_N$ was observed.
The STA was initially associated with BS~1 and the time at which the process started was not within a service disruption time, i.e., $j_N = 1$  and $c_N = 0$.
Thus, the initial state $s_N$ was set to $\bigl(x^{(1)}_{N},\dots, x^{(1)}_{1}, x^{(2)}_N, \dots, x^{(2)}_1, 1, 0\bigr)$.
The action $a_N$ was selected according to the $\epsilon$-greedy policy that is widely used in the learning phase in RL\cite{DQN2}; then, the next state $s_{N + 1}$ was set such that it contained the images $\bigl(x^{(1)}_{N + 1},\dots, x^{(1)}_{2}, x^{(2)}_{N + 1}, \dots, x^{(2)}_2\bigr)$, $j_{N + 1}$, and $c_{N + 1}$, where $j_{N + 1}$ and $c_{N + 1}$ were determined based on $a_N$ as discussed in Section~\ref{subsubsec:definition}.
The procedure was iterated and then ended when the state contained the last images $x^{(1)}_{T'}$ and $x^{(2)}_{T'}$.

The performance metric $R^{(j_{t + 1})}_{t + 1}$ for $j_{t + 1}
\in\mathcal{J}$ and $t\in\mathcal{T}_{1}$ in \eqref{eq:reward} was set as the capacity of the link between BS~$j_{t + 1}$ and the STA, and was calculated as follows:
The performance metric $R^{(j_{t + 1})}_{t + 1}$ was calculated by the Shannon capacity formula via the obtained received power value $P^{(j_{t + 1})}_{t + 1}$ as follows:
\begin{align*}
	R^{(j_{t + 1})}_{t + 1} = W_{j_{t + 1}}\log_{2}\!\left(1 + \frac{P^{(j_{t + 1})}_{t + 1}}{\sigma^2 W_{j_{t + 1}}}\right),
\end{align*}
where $\sigma^2$ and $W_{j_{t + 1}}$ denote the noise power spectral density and measurement bandwidth in the $j_{t + 1}$th measurement, respectively.

Subsequently, we evaluated the performance of the learned policy.
We simulated a decision process using the same procedure as the learning procedure with the exception that we used $\bigl(x^{(i)}_{t}\bigr)_{t\in\mathcal{T}_2}$ and $\bigl(P^{(i)}_{t}\bigr)_{t\in\mathcal{T}_2}$ for $i\in\{1, 2\}$, and the action was selected so that the learned optimal action-value in \eqref{eq:opt_q_func} is maximized.
We calculated the time average of the reward as a performance metric of the learned policy.

We iterated the learning and evaluation by using the same dataset.
We evaluated the policy that achieved the highest average reward throughout the iterations.
Parameters associated with the deep RL are summarized in Table~\ref{table:param}.

\begin{table}[t]
	\caption{Parameters Associated with RL}
	\label{table:param}
	\begin{center}
		\begin{tabular}{cc}	\toprule
			Discount factor, $\gamma$                                                           & 0.99                                    \\
			
			The number of obtained images, $T$                                                  & 16860                                   \\
			The number of images used for learning, $T'$                                                  & 11240  \\
			Number of iterations
			 of learning and evaluation                                                              & 1000                                    \\
			Exploration rate, $\epsilon$                                                         & 1--0.01 \\
			&(reduced by 0.01 \\
			&per iteration)\\
			Number of consecutive images, $N$                        & 2                                      \\
			Number of pixels in an input image, $P$                                             & $40\times 40$                           \\
			Interval between successive decision epochs $\tau$                                  & $30$\,ms                       \\
			Minibatch size\cite{DQN2}                                                                    & 32                                      \\
			Frequency of updating the target network\cite{DQN2}                                          & 10000                                 \\
			Noise power spectral density $\sigma^2$ & $-$173\,dBm/Hz\\
			\bottomrule
		\end{tabular}
	\end{center}
\end{table}

\subsection{Compared Framework}
We compared the proposed multi-camera operation with a single-camera operation.
To ensure fair comparison, we designed the decision process in a single-camera operation by formulating a decision process similar to that in Section~\ref{sec:formulation}---we replaced the images from multiple cameras in the definition of the state with the images from a single camera---and under this process, we learned a handover policy by deep RL.
In this decision process, the state at each decision epoch $t$ was set as follows:
\begin{align*}
	s_t = (x^{(1)}_t, \dots, x^{(1)}_{t - N + 1}, j_t, c_t).
\end{align*}

In this decision process, the network controller was blind to pedestrian~3.
Through the comparison, we validated that with the multi-camera operation, the blockage effects caused by pedestrian~3 were successfully predicted; thereby showing the feasibility of the multi-camera operation in the I2D-PH framework.


\subsection{Results}
\begin{figure}[t]
	\centering
	\subfigure[In multi-camera operation. Handover to BS~2 was triggered at 42.63\,s and handover to BS~1 was triggered at 42.90\,s.]{\includegraphics[width = 0.9\columnwidth]{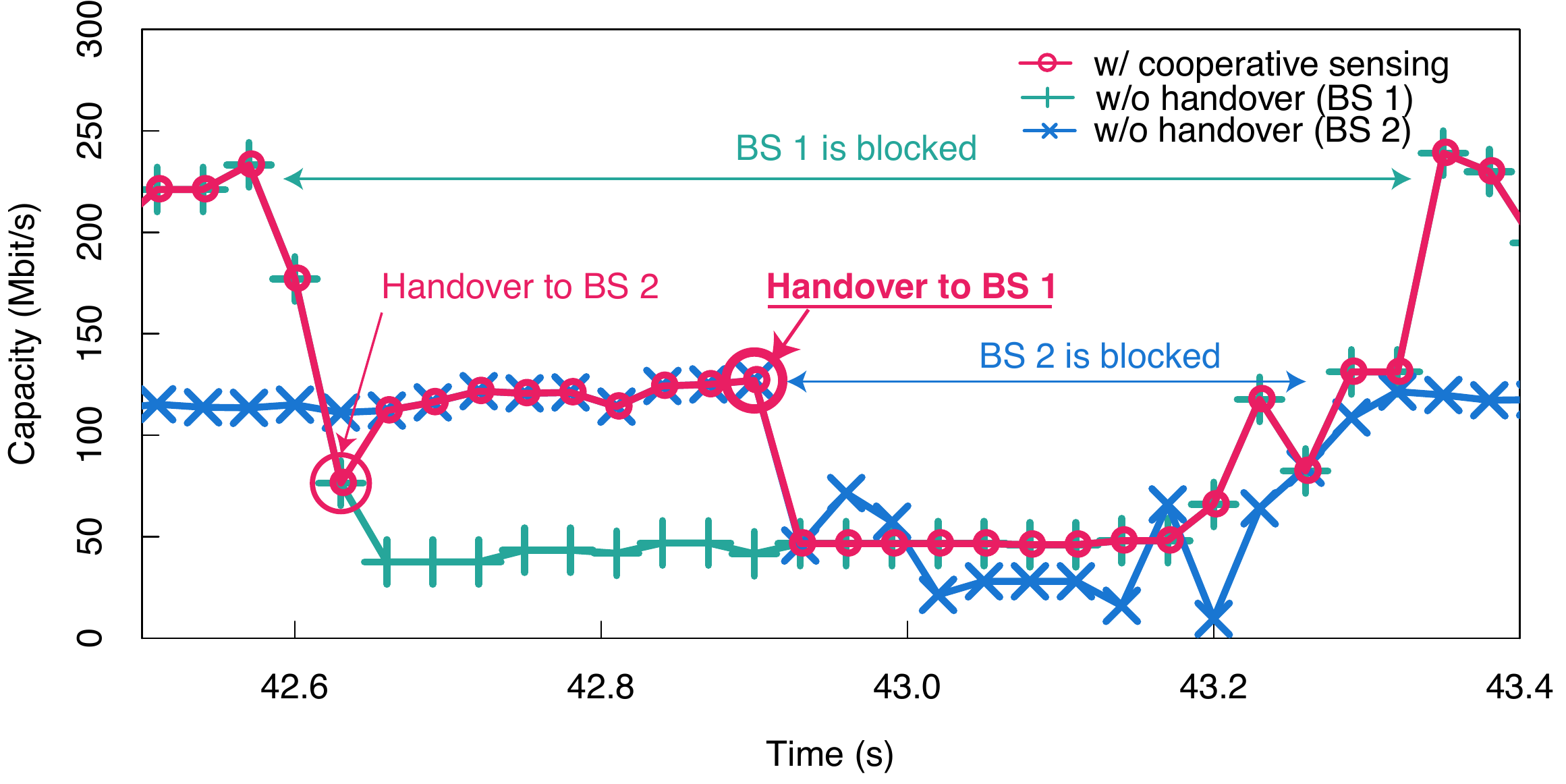}}
	\subfigure[In single-camera operation. Handover to BS~2 was triggered at 42.63\,s and handover to BS~1 was triggered at 43.20\,s.]{\includegraphics[width = 0.9\columnwidth]{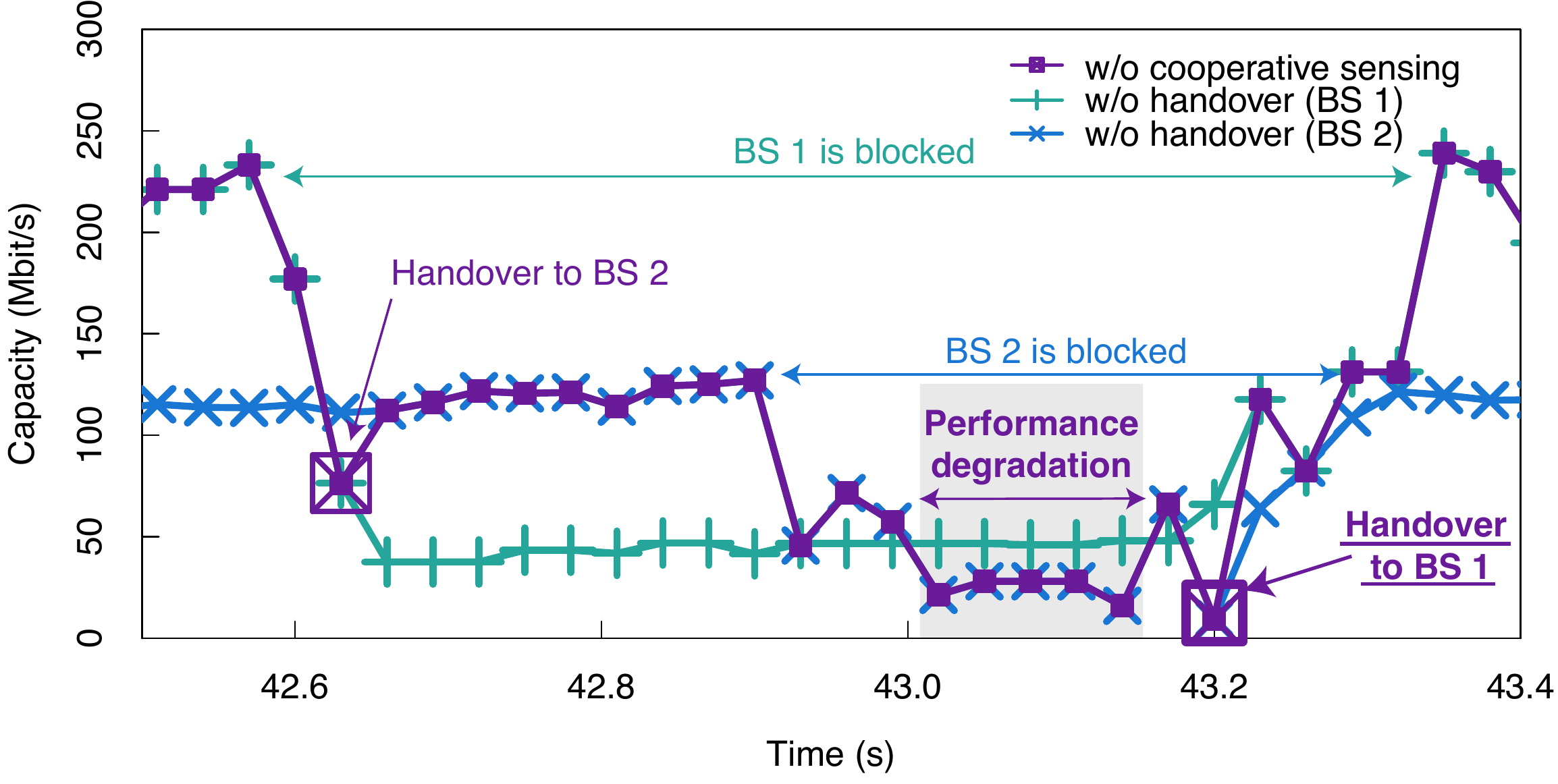}}
	\caption{Comparison between multi-camera operation and single-camera operation in terms of time series of capacity for service disruption time $T_{\mathrm{dis}} = 0$\,s.
	In the multi-camera operation, the blockage of the link between BS~2 and the STA could be predicted and a handover to BS~1 was triggered in a proactive fashion. In the single-camera operation, the handover to BS~1 was delayed because of the blindness to the blockage in BS~2.}
	\label{fig:data_rate_time_series}
\end{figure}

In Fig.~\ref{fig:data_rate_time_series}, we validate that camera~2 compensated for the blind spot of camera~1; thereby predicting blockage effects caused by the pedestrian~3 who is out of the coverage of camera~1.
Fig.~\ref{fig:data_rate_time_series} shows an example of the time series of the capacity for the service disruption time $T_{\mathrm{dis}} = 0$\,s.
The capacity of the link between BS~1 and the STA was degraded approximately from 42.6\,s to 43.3\,s and that of the link between BS~2 and the STA was degraded approximately from 42.9\,s to 43.4\,s.
Both frameworks successfully avoided the blockage effect in BS~1 by triggering a handover to BS~2\footnote{The handover to BS~2 was not triggered in a proactive fashion. This could be attributable to the fact that the capacity of the link between the STA and BS~2 was much lower than that between the the STA and BS~1. In the situation, a proactive handover did not contribute to the performance maximization; thus, the network controller triggered a handover in a reactive manner.}.
However, the framework without camera~2 remained to associate with BS~2 despite of the blockage in BS~2; hence experiencing greater degradation in the capacity relative to associating with BS~1.
Meanwhile, in the multi-camera operation, the network controller triggered a handover from BS~2 to BS~1 earlier and successfully avoided the degradation.

\begin{figure}[t]
	\centering
	\includegraphics[width = 0.75\columnwidth]{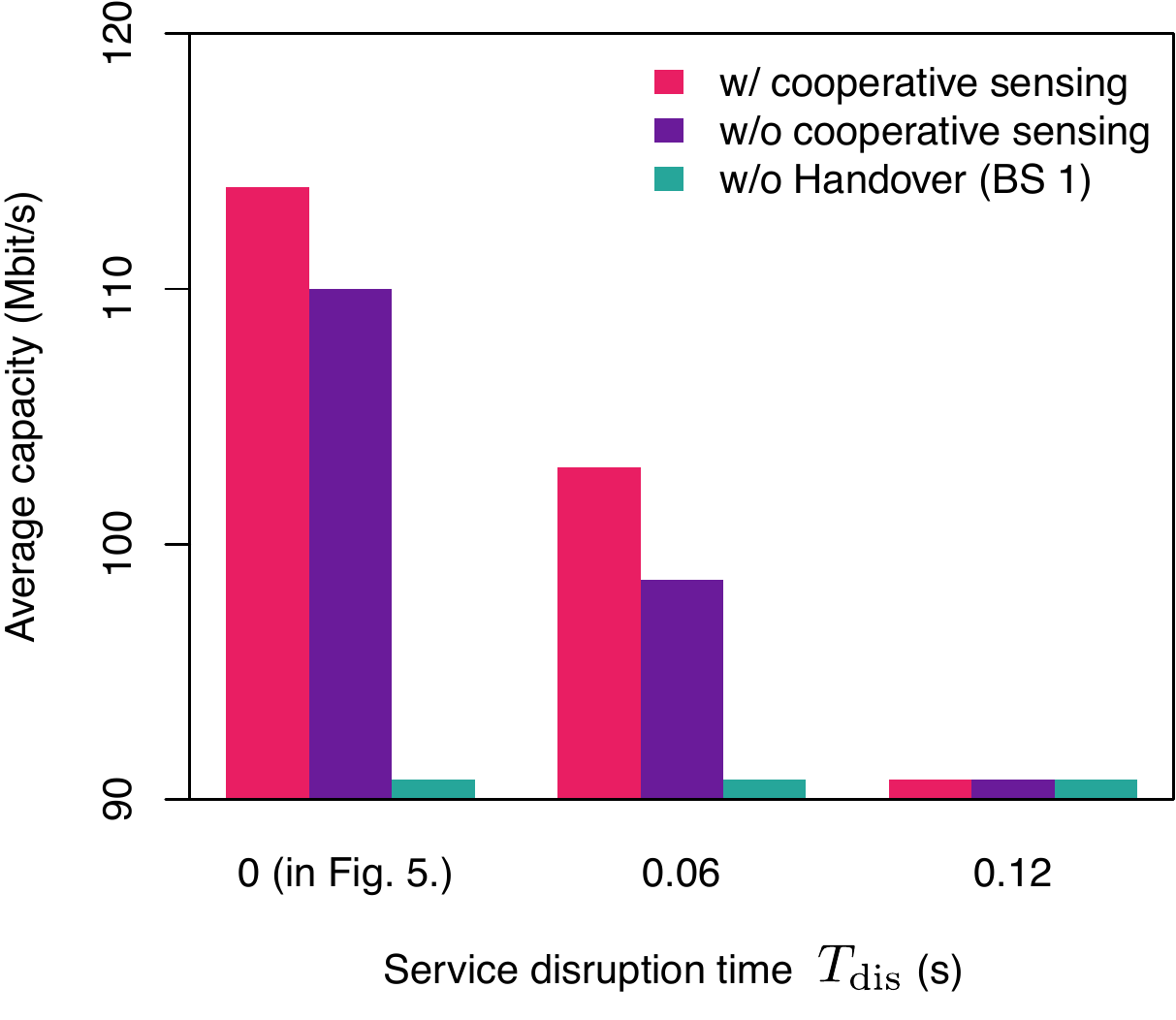}		
	\caption{Average capacity in Fig.~\ref{fig:data_rate_time_series} for different service disruption. In service disruption time $T_{\mathrm{dis}}=0, 0.06$\,s, the framework with multi-camera operation outperforms that with single-camera operation. In a larger service disruption, a handover is not triggered and the performance is equal to that without a handover.}
	\label{fig:average_reward}
\end{figure}

Fig.~\ref{fig:average_reward} confirms that the cooperative sensing with the multiple cameras contributed to performance enhancement.
Fig.~\ref{fig:average_reward} shows the capacity averaged over the duration plotted in Fig.~\ref{fig:data_rate_time_series} for different service disruption time $T_{\mathrm{dis}}$.
When $T_{\mathrm{dis}} = 0$\,s and $0.06$\,s, the framework with multi-camera operation outperformed the framework with single-camera operation.
The average capacity in the framework with the multi-camera operation was at most 5.10\% higher than that with the single-camera operation.

It should be notable that without explicitly estimating the positions/velocities of the pedestrians, we could learn the handover policy that led to performance enhancement in the multi-camera operation.
Thus, the results indicate the feasibility of adopting the cooperative sensing with multiple cameras to the I2D-PH framework.

Fig.~\ref{fig:average_reward} also shows that in a longer service disruption, i.e., when $T_{\mathrm{dis}} = 0.12$\,s, the average capacity in the both framework is equivalent to the average capacity in the link between the BS~1 and the STA.
The characteristic is attributed to the fact that the both frameworks do not trigger a handover because a handover does not contribute to the performance maximization.

\section{Conclusion}
\label{sec:conclusion}
In this paper, we proposed cooperative sensing with multi-camera operation in an I2D-PH framework for mmWave networks.
The framework directly maps camera images to a handover decision, thereby dealing with the situation where an arbitrary number of pedestrians cause blockage effects.
To successfully learn the optimal mapping in multi-camera operation, we designed a decision process where the state contains images from multiple cameras.
The evaluation based on the experimentally obtained camera images and received powers demonstrated the feasibility of learning the optimal mapping in the I2D-PH framework by showing that a camera successfully compensated for the blind spot of another camera.
The evaluation also showed that the framework with  multi-camera operation outperformed a conventional framework with single-camera operation.

\section*{Acknowledgment}
This work was supported in part by JSPS KAKENHI Grant Numbers JP17H03266 and JP18H01442, and by the KDDI Foundation.


\bibliographystyle{IEEEtran}
\bibliography{GLOBECOM2019}

\end{document}